\documentclass[prd,aps,twocolumn,a4paper,showkeys,nofootinbib]{revtex4-2}

\usepackage{amsmath}
\usepackage{amsfonts}
\usepackage{amssymb}	
\usepackage{graphicx}
\usepackage{amsbsy}
\usepackage{color}
\usepackage{commath}
\allowdisplaybreaks
\usepackage{hyperref}
\usepackage{makecell}
\usepackage{multirow}

\usepackage[normalem]{ulem}

\def\GMc2{G M_{\odot} c^{-2}}

\def\O{\mathcal{O}}

\def\lm{{\ell m}}

\def\lm{{\ell m}}

\def\lm{{\ell m}}

\def\l{{\ell }}

\def\O{{\cal O}}

\def\TEOBResumS{\texttt{TEOBResumS}}
\def\dali{\texttt{TEOBResumS\allowbreak-DALI}}

\newcommand\fnp[1]{{\hat{f}_{\varphi #1}^{\rm N _{nc}}}}
\def\NP22{{\fnp{,22}}}

\usepackage{color}
\definecolor{cyan}{rgb}{0,0.9,0.9}
\definecolor{orange}{rgb}{0.9,0.5,0}
\definecolor{magenta}{rgb}{1,0,1}
\definecolor{purple}{rgb}{0.8,0.4,0.8}
\definecolor{gray}{rgb}{0.8242,0.8242,0.8242}
\definecolor{dodgerblue}{rgb}{0.12, 0.56, 1.0}
\definecolor{darkgrey}{rgb}{0.5,0.5,0.5}
\definecolor{darkgreen}{rgb}{0,0.65,0}

\definecolor{colortab1}{rgb}{0.1, 0.1, 1.0}
\definecolor{colortab2}{rgb}{0.9,0,0.1}

\begin{document}
\title{2.5PN accurate waveform information for generic-planar-orbit binaries in
 effective one-body models}

\author{Andrea \surname{Placidi}${}^{1,2,3}$}
\author{Gianluca \surname{Grignani}${}^{1}$}
\author{Troels \surname{Harmark}${}^{2}$}
\author{Marta \surname{Orselli}${}^{1,2}$}
\author{Sara \surname{Gliorio}${}^{4}$}
\author{Alessandro \surname{Nagar}${}^{5,6}$}
\affiliation{${}^1$Dipartimento di Fisica e Geologia, Universit\`a di Perugia, INFN Sezione di Perugia, Via A. Pascoli, 06123 Perugia, Italia}
\affiliation{${}^2$Niels Bohr International Academy, Niels Bohr Institute, Copenhagen University,  Blegdamsvej 17, DK-2100 Copenhagen \O{}, Denmark}
\affiliation{${}^3$Galileo Galilei Institute for Theoretical Physics, Largo Enrico Fermi, 2, 50125 Firenze, Italy}
\affiliation{${}^4$ Gran Sasso Science Institute, Viale Francesco Crispi 7, 67100 L'Aquila, Italy}
\affiliation{${}^5$INFN Sezione di Torino, Via P. Giuria 1, 10125 Torino, Italy} 
\affiliation{${}^6$Institut des Hautes Etudes Scientifiques, 91440 Bures-sur-Yvette, France}

\begin{abstract}
	We provide the post-Newtonian (PN) waveform for binary systems in motion along generic planar orbits at 2.5PN accuracy, in terms of the dynamical variables of the effective one-body (EOB) formalism. In addition to the calculation of the higher order terms for all the contributions to the waveform that have been already considered in previous avatars of EOB models, we also compute the EOB expression of the oscillatory memory terms. These are purely non-circular contributions, first appearing at 1.5PN order, that have been so far neglected in the EOB literature. This should foster their inclusion in EOB models and the definitive assessment of their role in shaping gravitational wave signals at infinity. To further promote the application of our results, we also derive associated non-circular factors according to the waveform factorization prescription of the non-circular EOB model \dali{}; the result is a set of ready-to-use non-circular factors that can be directly implemented as extra non-circular corrections in the waveform of \dali{}.
\end{abstract}
\date{\today}

\maketitle

\section{Introduction}
\label{sec:introduction}
From the observation of the first gravitational wave (GW) signal~\cite{Abbott:2016blz}, the LIGO-Virgo-Kagra (LVK) collaboration~\cite{LIGOScientific:2021djp} has been bringing prestige to the field of gravitational wave astronomy with numerous confirmed detections \cite{LIGOScientific:2018mvr,Abbott:2020niy,LIGOScientific:2021psn}. Prompted by these results, the scientific community is now very active in improving the already existing GW interferometers and in setting up the commissioning for new detectors, both ground based such as Einstein Telescope~\cite{Maggiore:2019uih} and Cosmic Explorer~\cite{Evans:2021gyd}, or designed to work in space such as  LISA~\cite{LISA:2017pwj} and TianQin~\cite{Mei:2021boh}.

The prominent sources of the GW signals observed by current and future detectors are compact binary coalescences (CBC), i.e.~binary systems made by black holes and neutron stars. 
Therefore, the data we can gather from GW astronomy offer an unprecedented chance to investigate the properties of these compelling systems and probe General Relativity 
in the strong field regime. The extrapolation and analysis of these data is heavily based on huge banks of GW templates, which must comprehensively cover the 
space of the relevant CBC parameters and be accurate enough to sustain data analysis, with specific requirements depending on the sensitivity of the given detector~\cite{Purrer:2019jcp}. 

Amid the different directions of development for such CBC waveform models, both the analytical and numerical relativity communities have recently turned their focus on extending their CBC waveforms from their native quasi-circular-orbit implementation to more generic, {\it non-circularized} binary dynamics~\cite{Hinder:2017sxy,Hinderer:2017jcs,Chiaramello:2020ehz,Nagar:2020xsk,Islam:2021mha,Nagar:2021gss,Nagar:2021xnh,Albanesi:2021rby,Liu:2021pkr,Yun:2021jnh,Tucker:2021mvo,Setyawati:2021gom,Khalil:2021txt,Ramos-Buades:2021adz,Placidi:2021rkh,Albanesi:2022xge,Albanesi:2022ywx}. This has been mainly spurred by the increasing observational relevance that non-circularized binaries have been gaining over the past few years~\cite{LIGOScientific:2020iuh,Gayathri:2020coq,Gamba:2021gap}, and by the many GW detections from non-circular CBC that are expected in view of the forthcoming next generation of interferometers, LISA in particular~\cite{LISA:2017pwj,Babak:2017tow,Gair:2017ynp}.

Specializing the discussion to effective-one-body (EOB) models~\cite{Buonanno:1998gg,Buonanno:2000ef,Damour:2000we,Damour:2001tu,Damour:2015isa}, which combine analytical and numerical information into a unified, robust, and theoretically comprehensive description of the whole coalescence process, the generalization to non-circular binaries is proceeding according to different strategies of implementation: on the one hand, in the eccentric branch of~\TEOBResumS{} now known as \dali{}\cite{Nagar:2020pcj,Nagar:2021gss,Riemenschneider:2021ppj,Nagar:2021xnh}, it is basically realized by replacing the quasi-circular expression of the Newtonian prefactor, in the factorization of the spherical modes of the waveform, with the general expression obtained by computing the time-derivatives of the Newtonian mass and current multipoles~\cite{Chiaramello:2020ehz,Nagar:2020xsk,Nagar:2021gss,Nagar:2021xnh}; on the other hand, the eccentric model {\tt SEOBNRv4HME}~\cite{Ramos-Buades:2021adz} has been built around {\it post-Newtonian} (PN) waveform results for generic-planar-orbit dynamics, previously released in a series of works \cite{Mishra:2015bqa,Khalil:2021txt,Boetzel:2019nfw,Ebersold:2019kdc}. More specifically, this waveform information has been implemented up to next-to-next-to-leading-order in the PN expansion, i.e.~at 2PN.\footnote{We recall in this respect that the PN expansion is essentially an expansion for small internal velocities of the considered system, and it is usually organized in powers of $1/c$, with $c$ being the speed of light. In particular, terms proportional to $1/c^n$ correspond to corrections at $\frac{n}{2}$PN order.}  Subsequently, this same PN information has been also incorporated into \dali{}; see Ref.~\cite{Placidi:2021rkh}. Here, after suitable factorization and resummation procedures, the extra analytical information has been recast in the form of extra waveform factors, which were observed to improve, slightly but noticeably, the analytical/numerical phase agreement of the model, although with a marginal impact on the waveform amplitude, which was seen to benefit mostly from the generic Newtonian prefactor prescription.
Other significant differences between the aforementioned EOB models can be found at the level of the radiation reaction forces, for which we refer to the thorough analysis of Ref.~\cite{Albanesi:2022ywx}. 

Moreover, Ref.~\cite{Albanesi:2022xge} has recently proposed a novel approach to the inclusion of non-circular PN information in the {\it instantaneous} (i.e.~not involving time integrals over the past history of the source) part of the EOB waveform. The main idea behind this proposal consists in \emph{not} performing the usual order-reduction procedure, by which the natural occurring time-derivatives in the waveform are replaced with the respective PN-expanded equations of motion (EOM); instead, all these time derivatives are left explicit and evaluated along the dynamics, as it is done in the generic Newtonian prefactor introduced in Ref.~\cite{Chiaramello:2020ehz}. As a consequence, each time derivative is \emph{de facto} evaluated using either the \emph{exact} or the full \emph{EOB-resummed} EOM, depending on whether one has to deal, respectively, with the test-mass or the comparable mass scenarios. In both cases, this new strategy has been proven to further improve the non-circular correction to the waveform amplitude, which no longer vanishes at the turning points of the radial motion, the apastra and periastra of the orbits along which the binary system evolves; see e.g.~Fig.~2 in Ref.~\cite{Albanesi:2022xge}. A related improvement has been also observed in the analytical/numerical agreement of the model at the level of the corresponding fluxes of energy and angular momentum at infinity; see Fig.~4 in Ref.~\cite{Albanesi:2022xge}.

Following this line of work, in this paper we push the computation of non-circular waveform information in the EOB formalism up to the 2.5PN order, encompassing also the so far neglected oscillatory memory terms; then, we use our findings to derive non-circular waveform factors that can be used to upgrade the PN sector of the generic-orbit model \dali{}.

More in detail, the generic-orbit waveform contributions we provide here, for the first time, in EOB coordinates, are
\begin{itemize}
	\item[(i)] the 2.5PN tail terms.
	\item[(ii)] the 2.5PN post-adiabatic term.
	\item[(iii)] the 1.5PN, 2PN, and 2.5PN oscillatory memory terms.
	\item[(iv)] the 2.5PN instantaneous terms with implicit time-derivatives.
\end{itemize}
The contributions (i)-(iii) are obtained by expressing in EOB coordinates the results of Refs.~\cite{Boetzel:2019nfw,Ebersold:2019kdc}, where the authors completed the derivation of the full 3PN waveform for non-circular binaries, adopting the quasi-Keplerian parametrization in harmonic coordinates. For the contribution (iv) instead we push at 2.5PN the scheme proposed and outlined in Ref.~\cite{Albanesi:2022xge}. The expressions of the new waveform contributions for the specific case of the dominant $\l=m=2$ mode are given in Eq.~\eqref{eq:inst2.5PN} and Eqs.~\eqref{eob_leading_osc_mem}-\eqref{eq:post_ad_22}; all the other spherical modes can be found in the file \texttt{2.5PN\_results\_in\_EOB\_ coordinates.m}, provided as supplementary material to this paper. 

Regarding the details of the EOB dynamics, we refer the reader to Refs.~\cite{Chiaramello:2020ehz,Nagar:2021gss}.

The paper is organized as follows. In Sec.~\ref{sec:waveform} we go through the derivation of the various terms which compose the 2.5PN waveform in EOB coordinates, dealing separately with the adiabatic instantaneous terms in Sec.~\ref{sec:InstNovVague} and with all the others in Sec.~\ref{sec:EOM_terms}. These results are then used in Sec.~\ref{sec:nc_factors} to derive 2.5PN-accurate  non-circular multiplicative corrections to the factorized waveform modes of \dali{}. Finally, we summarize and comment our results in Sec.~\ref{sec:Conclusions}.
\section{2.5PN generic-orbit waveform in EOB coordinates}
\label{sec:waveform}
We consider a non-spinning black hole binary system with individual masses $m_{1,2}$. It is useful to define the total mass $M \equiv m_1+m_2$, the reduced mass $\mu\equiv m_1 m_2/M$ and the symmetric mass ratio $\nu\equiv \mu/M$. The gravitational wave strain $h$ at future null infinity can be decomposed as
\begin{equation}
	\label{h_spherical_dec}
	h\equiv h_+ - i h_\times = D_L^{-1} \sum_{\ell=2}^{+\infty} \sum_{m=-\ell}^{\ell} h_{\ell m} \, {}_{-2}Y_{\ell m},
\end{equation}
where $D_L$ is the luminosity distance of the source and ${}_{-2}Y_{\ell m}$ 
are the spin-weight $-2$ spherical harmonics. In what follows we discuss the 2.5PN-accurate derivation in EOB coordinates of the spherical multipoles $h_{\ell m}$, limiting ourselves to the $m\neq0$ case. We employ the usual mass-reduced EOB phase-space variables $(r, \varphi,p_{r_*},p_\varphi)$, which are given by: $r\equiv R/M $, the relative separation in the center of mass frame; $\varphi$, the orbital phase; $p_{r_*} \equiv (A/B)^{1/2} \, p_r$, where $A,B$ are 
the radial potentials entering the EOB effective metric \cite{Chiaramello:2020ehz,Nagar:2021gss} and $p_r\equiv P_R/\mu$ is the radial momentum; $p_\varphi\equiv P_\varphi / \mu M$, the angular momentum.

In general, the PN expression of each spherical waveform mode consists of two distinct parts: the \emph{instantaneous} part, which depends on the state of the source at a specific retarded time, and the \emph{hereditary} part, whose time dependence is instead extended to the past history of the source via distinctive time integrals. Among the hereditary contributions there is a further distinction based on the traits of their time dependence: the \emph{tail} terms are those that are progressively suppressed as one goes towards the remote past of the source, while the \emph{memory} terms weight equally each moment of the source past.

The derivation of instantaneous and hereditary waveform components will proceed according to different approaches, respectively outlined in Sec.~\ref{sec:InstNovVague} and Sec.~\ref{sec:EOM_terms}. We specify however that in the latter section we will also target the additional phasing terms of instantaneous type that are induced by the energy and angular momentum loss, the so called \emph{post-adiabatic} terms \cite{Boetzel:2019nfw}, for which the approach of Sec.~\ref{sec:InstNovVague} is not viable; we anticipate that terms of this kind arise as 2.5PN contributions in the PN expansion of each spherical mode and, correspondingly, only those of the $\ell=m=2$ mode are relevant for the 2.5PN accuracy in the total waveform \eqref{h_spherical_dec} we are aiming for here.

\subsection{Time-derivative dependent instantaneous part}
\label{sec:InstNovVague}
The instantaneous contributions to the waveform are dealt with according to the strategy introduced in Ref.~\cite{Albanesi:2022xge}, which remarkably improves the behavior of the waveform at the apastra and periastra of the binary motion. In general, for non-precessing binaries, the spherical multipoles of the waveform in the decomposition~\eqref{h_spherical_dec} are given by
\begin{align}
	\label{hlm_to_UV}
	h_{\lm} &=-\frac{ U_{\lm}}{\sqrt{2}c^{\ell+2}}  \quad \quad&\text{if $\ell+m$ is even}, \cr
	h_{\lm} &=i \frac{ V_{\lm}}{\sqrt{2}c^{\ell+3}}  &\text{if $\ell+m$ is odd},
\end{align}
in terms of the mass-type ($U_{\lm}$) and current-type ($V_{\lm}$) radiative multipole moments. The latter are related to the symmetric trace-free (STF) radiative moments $U_L$ and $V_L$ by
\begin{align}
	\label{UVlm_to_UVL}
	U_{\lm} &= \frac{4}{\ell!}\sqrt{\dfrac{(\ell+1)(\ell+2)}{2\ell(\ell+1)}} \alpha^L_{\lm} U_L,\cr
	V_{\lm} &= -\frac{8}{\ell!}\sqrt{\dfrac{\ell(\ell+2)}{2(\ell+1)(\ell-1)}} \alpha^L_{\lm} V_L,\cr
\end{align}
where $L\equiv i_1, \ldots ,i_\ell$ and $\alpha^L_{\lm}$ are the STF tensors which connect the basis of spherical harmonics to the set of STF tensors $N_{\langle L \rangle} \equiv N_{\langle i_1} \cdots N_{i_\ell \rangle}$,\footnote{Adopting a standard notation the angular brackets $\langle\rangle$ denote a STF projection over the indices they enclose. } defined by the unit vector $\boldsymbol{N}$ pointing from the source of the wave to its observer. The STF radiative moments $(U_L,V_L)$ are computed in terms of the STF moments of the source $(I_L,J_L,X_L,Y_L,Z_L,W_L)$\footnote{Here $I_L$ are the mass-type source moments, $J_L$ the current-type ones, and all the others are typically dubbed gauge moments since they are of gauge nature in linearized gravity, even though they are physical in the full nonlinear theory. } following the PN-matched multipolar post-Minkowskian scheme developed by Blanchet and Damour \cite{Blanchet:1989ki} and  reviewed in Ref.~\cite{Blanchet:2013haa}. The relations connecting the two set of STF moments are conveniently collected, at 3PN accuracy, in Sec.~IIIA of Ref.~\cite{Mishra:2015bqa}. Moreover Sec.~IIIB of Ref.~\cite{Mishra:2015bqa} collects the  expression at 3PN-order for the source moments of nonspinning binary systems in harmonic coordinates. What is of uttermost importance for our approach is the presence of time derivatives in the aforementioned relations between radiative and source moments. For instance at the level of the mass quadrupole, which is needed for the $\ell=m=2$ spherical mode, and focusing just on the instantaneous terms, we have
\begin{align}
	\label{radiative_mass_quadrupole}
	U^{\rm inst}_{ij}&= I_{ij}^{(2)}+\frac{G}{c^5} \bigg[\frac{1}{7}I_{a \langle i}^{(5)}I_{j \rangle a}-\frac{5}{7}I_{a \langle i}^{(4)}I_{j \rangle a}^{(1)}-\frac{2}{7}I_{a \langle i}^{(3)}I_{j \rangle a}^{(2)}\cr
	& + \frac{1}{3} \varepsilon_{ab \langle i}I_{j \rangle a}^{(4)}J_b + 4\bigg(W^{(4)}I_{ij}+W^{(3)}I_{ij}^{(1)}-W^{(2)}I_{ij}^{(2)}\cr
	& -W^{(1)}I_{ij}^{(3)}\bigg) \bigg]+\mathcal{O}(c^{-6}),
\end{align}
where the superscript $(n)$ indicates the $n$th time derivative.
In the standard approach, also followed in Ref.~\cite{Mishra:2015bqa}, all the time derivatives in this expressions are removed with an order-reduction procedure which replaces them with the corresponding PN-expanded EOM, truncated at the desired PN accuracy. On the contrary, embracing the strategy proposed in Ref.~\cite{Albanesi:2022xge}, we choose \emph{not} to remove the time derivatives in the instantaneous terms, and instead keep them in explicit form. This amounts by all means to a PN generalization of what has been done in Ref.~\cite{Chiaramello:2020ehz} to define the generic-orbit  Newtonian prefactor, which in fact turns out to be precisely the leading order term of the instantaneous waveform component we compute here.

The procedure to derive these results in EOB coordinates is the following: starting from the harmonic coordinate source moments, given in Sec.~IIIB of Ref.~\cite{Mishra:2015bqa}, we recast them in EOB coordinates with the transformations given in Eqs.~(5)-(8) of Ref.~\cite{Placidi:2021rkh}.\footnote{Tough 2PN-accurate, these transformations can still be used for our 2.5PN computation since their next PN contribution would appear at 3PN.} Then we insert the so obtained source moments, now functions of $(r, \varphi,p_{r_*},p_\varphi)$, in the relations connecting them to the radiative moments, such as Eq.~\eqref{radiative_mass_quadrupole}. In all these relations one has time-derivatives of the source moments. Therefore, when they are used together with Eqs.~\eqref{hlm_to_UV} and \eqref{UVlm_to_UVL} to compute the instantaneous part of the waveform, the latter shows a dependence on several time derivatives of the EOB variables. For instance, the formal structure of the instantaneous terms in $h_{22}$, up to 2.5PN, read
\begin{align}
	\label{h22_inst}
	h_{22}^{\rm inst} &= h_{22}^{N}\left(r^{(2\uparrow)},\Omega^{(1\uparrow)}\right)\cr
	&+\frac{1}{c^2}h_{22}^{\rm 1PN_{\rm inst}}\left(r^{(2\uparrow)},\Omega^{(1\uparrow)},p_{r_*}^{(2\uparrow)},p_{\varphi}^{(2\uparrow)}\right)\cr
	&+\frac{1}{c^4}h_{22}^{\rm 2PN_{\rm inst}}\left(r^{(2\uparrow)},\Omega^{(1\uparrow)},p_{r_*}^{(2\uparrow)},p_{\varphi}^{(2\uparrow)}\right)\cr
	&+\frac{1}{c^5}h_{22}^{\rm 2.5PN_{\rm inst}}\left(r^{(5\uparrow)},\Omega^{(4\uparrow)},p_{r_*}^{(4\uparrow)},p_{\varphi}^{(2\uparrow)}\right),
\end{align}
where $\Omega\equiv\dot{\varphi}$ and the symbol ${(n \! \! \uparrow)}$ indicates that the corresponding variable appears with all its time derivatives up to the $n$th. As anticipated before, the leading term of Eq.~\eqref{h22_inst} coincides with the generic $\ell=m=2$ Newtonian prefactor \cite{Chiaramello:2020ehz}
\begin{align}
	\label{eq:Newt_prefactor}
	h_{22}^{N} & = -\frac{8 \nu}{c^4}\sqrt{\dfrac{\pi}{5}} \, r^2 \Omega^2 e^{-2{\rm i}\varphi}
	\hat{h}_{22}^{N_{\rm nc}}, \\
	\hat{h}_{22}^{N_{\rm nc}} & =1-\dfrac{1}{2}\left(\dfrac{\dot{r}^2}{r^2\Omega^2} 
	+ \dfrac{\ddot{r}}{r\Omega^2}\right) 
	+ {\rm i}\left(\dfrac{2\dot{r}}{r\Omega} + \dfrac{\dot{\Omega}}{2\Omega^2}\right)
\end{align}
where the subscript ``${\rm nc}$" labels waveform contributions that are purely non-circular, i.e.~that vanish when the dynamics of the source is constrained on circular orbits. The remaining PN terms are given by
\begin{widetext}
\begin{align}
    h_{22}^{\rm 1PN_{\rm inst}} & = \frac{2}{21} \sqrt{\frac{\pi }{5}} \nu  e^{-2 i \varphi } \bigg[ (90 \nu +54) p_{r_*}^2 r^2 \Omega ^2+9 i (5 \nu +3) p_{r_*}^2 r^2 \dot{\Omega}-9 (5 \nu +3) p_{r_*}^2 r \ddot{r}+36 i (5 \nu +3) p_{r_*}^2 r \dot{r} \Omega \cr&-9 (5 \nu +3) p_{r_*}^2 \dot{r}^2-9 (5 \nu +3) p_{r_*} \ddot{p}_{r_*} r^2+36 i (5 \nu +3) p_{r_*} \dot{p}_{r_*} r^2 \Omega -36 (5 \nu +3) p_{r_*} \dot{p}_{r_*} r \dot{r}-8 i (6 \nu +5) p_{r_*} p_\varphi r \Omega ^2\cr&+4 (6 \nu +5) p_{r_*} p_\varphi r \dot{\Omega}+2 i (6 \nu +5) p_{r_*} p_\varphi \ddot{r}+8 (6 \nu +5) p_{r_*} p_\varphi \dot{r} \Omega +2 i (6 \nu +5) p_{r_*} \ddot{p}_\varphi r+8 (6 \nu +5) p_{r_*} \dot{p}_\varphi r \Omega \cr& +4 i (6 \nu +5) p_{r_*} \dot{p}_\varphi \dot{r}+2 i (6 \nu +5) \ddot{p}_{r_*} p_\varphi r-9 (5 \nu +3) \dot{p}_{r_*}^2 r^2+8 (6 \nu +5) \dot{p}_{r_*} p_\varphi r \Omega +4 i (6 \nu +5) \dot{p}_{r_*} p_\varphi \dot{r}\cr&+4 i (6 \nu +5) \dot{p}_{r_*} \dot{p}_\varphi r+(42 \nu +14) p_\varphi^2 \Omega ^2+7 i (3 \nu +1) p_\varphi^2 \dot{\Omega}-7 (3 \nu +1) p_\varphi \ddot{p}_\varphi+28 i (3 \nu +1) p_\varphi \dot{p}_\varphi \Omega \cr&-7 (3 \nu +1) \dot{p}_\varphi^2+12 (\nu -19) r \Omega ^2+6 i (\nu -19) r \dot{\Omega}+(57-3 \nu ) \ddot{r}+12 i (\nu -19) \dot{r} \Omega \bigg] \\
    h_{22}^{\rm 2PN_{\rm inst}} & = \frac{1}{378 r^4}\sqrt{\frac{\pi }{5}} \nu  e^{-2 i \varphi } \bigg\{ -90 p_{r_*}^4 [(\nu -13) \nu +5] \Omega ^2 r^6+270 p_{r_*}^2 \dot{p}_{r_*}^2 [(\nu -13) \nu +5] r^6+90 p_{r_*}^3 \ddot{p}_{r_*} [(\nu -13) \nu +5] r^6\cr&-360 i p_{r_*}^3 \dot{p}_{r_*} [(\nu -13) \nu +5] \Omega  r^6-45 i p_{r_*}^4 [(\nu -13) \nu +5] \dot{\Omega} r^6-48 i p_{r_*}^3 p_\varphi [\nu  (14 \nu +43)-5] \Omega ^2 r^5\cr&+24 p_{r_*}^2 [\nu  (47 \nu +91)+124] \Omega ^2 r^5+45 p_{r_*}^4 \ddot{r} [(\nu -13) \nu +5] r^5+360 p_{r_*}^3 \dot{p}_{r_*} \dot{r} [(\nu -13) \nu +5] r^5\cr&+72 i p_{r_*} \dot{p}_{r_*}^2 p_\varphi [\nu  (14 \nu +43)-5] r^5+36 i p_{r_*}^2 \ddot{p}_{r_*} p_\varphi [\nu  (14 \nu +43)-5] r^5+12 i p_{r_*}^3 \ddot{p}_\varphi [\nu  (14 \nu +43)-5] r^5\cr&+72 i p_{r_*}^2 \dot{p}_{r_*} \dot{p}_\varphi [\nu  (14 \nu +43)-5] r^5-12 \dot{p}_{r_*}^2 [\nu  (47 \nu +91)+124] r^5-12 p_{r_*} \ddot{p}_{r_*} [\nu  (47 \nu +91)+124] r^5\cr&-180 i p_{r_*}^4 \dot{r} [(\nu -13) \nu +5] \Omega  r^5+144 p_{r_*}^2 \dot{p}_{r_*} p_\varphi [\nu  (14 \nu +43)-5] \Omega  r^5+48 p_{r_*}^3 \dot{p}_\varphi [\nu  (14 \nu +43)-5]\Omega  r^5\cr&+48 i p_{r_*} \dot{p}_{r_*} [\nu  (47 \nu +91)+124] \Omega  r^5+24 p_{r_*}^3 p_\varphi [\nu  (14 \nu +43)-5] \dot{\Omega} r^5+12 i p_{r_*}^2 [\nu  (47 \nu +91)+124] \dot{\Omega} r^5\cr&+36 p_{r_*}^2 p_\varphi^2 (-21 \nu ^2+\nu -25) \Omega ^2 r^4-8 i p_{r_*} p_\varphi [4 \nu  (22 \nu +353)+983] \Omega ^2 r^4+[24 \nu  (65 \nu +271)+924] \Omega ^2 r^4\cr&+45 p_{r_*}^4 \dot{r}^2 [(\nu -13) \nu +5] r^4+12 i p_{r_*}^3 p_\varphi \ddot{r} [\nu  (14 \nu +43)-5] r^4+72 i p_{r_*}^2 \dot{p}_{r_*} p_\varphi \dot{r} [\nu  (14 \nu +43)-5] r^4\cr&+24 i p_{r_*}^3 \dot{p}_\varphi \dot{r} [\nu  (14 \nu +43)-5] r^4+18 \dot{p}_{r_*}^2 p_\varphi^2 [\nu  (21 \nu -1)+25] r^4+18 p_{r_*} \ddot{p}_{r_*} p_\varphi^2 [\nu  (21 \nu -1)+25] r^4\cr&+18 p_{r_*}^2 \dot{p}_\varphi^2 [\nu  (21 \nu -1)+25] r^4+18 p_{r_*}^2 p_\varphi \ddot{p}_\varphi [\nu  (21 \nu -1)+25] r^4+72 p_{r_*} \dot{p}_{r_*} p_\varphi \dot{p}_\varphi [\nu  (21 \nu -1)+25] r^4\cr&+2 i \ddot{p}_{r_*} p_\varphi [4 \nu  (22 \nu +353)+983] r^4+2 i p_{r_*} \ddot{p}_\varphi (4 \nu  (22 \nu +353)+983) r^4+4 i \dot{p}_{r_*} \dot{p}_\varphi [4 \nu  (22 \nu +353)+983] r^4\cr&-6 p_{r_*}^2 \ddot{r} [\nu  (47 \nu +91)+124] r^4-24 p_{r_*} \dot{p}_{r_*} \dot{r} [\nu  (47 \nu +91)+124] r^4+48 p_{r_*}^3 p_\varphi \dot{r} [\nu  (14 \nu +43)-5] \Omega  r^4\cr&-72 i p_{r_*} \dot{p}_{r_*} p_\varphi^2 [\nu  (21 \nu -1)+25] \Omega  r^4-72 i p_{r_*}^2 p_\varphi \dot{p}_\varphi [\nu  (21 \nu -1)+25] \Omega  r^4+8 \dot{p}_{r_*} p_\varphi (4 \nu  [22 \nu +353)+983] \Omega  r^4\cr&+8 p_{r_*} \dot{p}_\varphi [4 \nu  (22 \nu +353)+983] \Omega  r^4+24 i p_{r_*}^2 \dot{r} [\nu  (47 \nu +91)+124] \Omega  r^4-18 i p_{r_*}^2 p_\varphi^2 [\nu  (21 \nu -1)+25] \dot{\Omega} r^4\cr&+4 p_{r_*} p_\varphi [4 \nu  (22 \nu +353)+983] \dot{\Omega} r^4+6 i [2 \nu  (65 \nu +271)+77] \dot{\Omega} r^4+4 p_\varphi^2 (434 \nu ^2-3806 \nu -1703) \Omega ^2 r^3\cr&-144 i p_{r_*} p_\varphi^3 [\nu  (6 \nu +11)-5] \Omega ^2 r^3+\dot{p}_\varphi^2 (-868 \nu ^2+7612 \nu +3406) r^3+p_\varphi \ddot{p}_\varphi (-868 \nu ^2+7612 \nu +3406) r^3\cr&+36 i \ddot{p}_{r_*} p_\varphi^3 [\nu  (6 \nu +11)-5] r^3+216 i p_{r_*} p_\varphi \dot{p}_\varphi^2 (\nu  [6 \nu +11)-5] r^3+108 i p_{r_*} p_\varphi^2 \ddot{p}_\varphi [\nu  (6 \nu +11)-5] r^3\cr&+216 i \dot{p}_{r_*} p_\varphi^2 \dot{p}_\varphi [\nu  (6 \nu +11)-5] r^3+8 i p_\varphi \dot{p}_\varphi (434 \nu ^2-3806 \nu -1703) \Omega  r^3+144 \dot{p}_{r_*} p_\varphi^3 [\nu  (6 \nu +11)-5] \Omega  r^3\cr&+432 p_{r_*} p_\varphi^2 \dot{p}_\varphi [\nu  (6 \nu +11)-5] \Omega  r^3+2 i p_\varphi^2 (434 \nu ^2-3806 \nu -1703) \dot{\Omega} r^3+72 p_{r_*} p_\varphi^3 [\nu  (6 \nu +11)-5] \dot{\Omega} r^3\cr&+p_\varphi^4 [30-6 \nu  (143 \nu +109)] \Omega ^2 r^2+p_\varphi^2 \ddot{r} (434 \nu ^2-3806 \nu -1703) r^2+4 p_\varphi \dot{p}_\varphi \dot{r} (434 \nu ^2-3806 \nu -1703) r^2\cr&+p_\varphi^3 \ddot{p}_\varphi (858 \nu ^2+654 \nu -30) r^2-36 i p_{r_*} p_\varphi^3 \ddot{r} [\nu  (6 \nu +11)-5] r^2-72 i \dot{p}_{r_*} p_\varphi^3 \dot{r} [\nu  (6 \nu +11)-5] r^2\cr&-216 i p_{r_*} p_\varphi^2 \dot{p}_\varphi \dot{r} [\nu  (6 \nu +11)-5] r^2+18 p_\varphi^2 \dot{p}_\varphi^2 [\nu  (143 \nu +109)-5] r^2-4 i p_\varphi^2 \dot{r} (434 \nu ^2-3806 \nu -1703) \Omega  r^2\cr&-144 p_{r_*} p_\varphi^3 \dot{r} [\nu  (6 \nu +11)-5] \Omega  r^2-24 i p_\varphi^3 \dot{p}_\varphi [\nu  (143 \nu +109)-5] \Omega  r^2-3 i p_\varphi^4 [\nu  (143 \nu +109)-5] \dot{\Omega} r^2\cr&+p_\varphi^2 \dot{r}^2 (-868 \nu ^2+7612 \nu +3406) r+72 i p_{r_*} p_\varphi^3 \dot{r}^2 [\nu  (6 \nu +11)-5] r-24 p_\varphi^3 \dot{p}_\varphi \dot{r} [\nu  (143 \nu +109)-5] r\cr&+p_\varphi^4 \ddot{r} [15-3 \nu  (143 \nu +109)] r+12 i p_\varphi^4 \dot{r} [\nu  (143 \nu +109)-5] \Omega  r+9 p_\varphi^4 \dot{r}^2 [\nu  (143 \nu +109)-5] \bigg\} \\
    \label{eq:inst2.5PN}
    h_{22}^{\rm 2.5PN_{\rm inst}} & = \frac{2 i }{21 r^3}\sqrt{\frac{\pi }{5}} \nu ^2 e^{-2 i \varphi } \bigg( 16 \Omega ^5 r^7-60 \Omega  \dot{\Omega}^2 r^7-40 \Omega ^2 \ddot{\Omega} r^7-10 i \Omega  \Omega^{(3)} r^7+\Omega^{(4)} r^7+80 i \Omega ^3 \dot{\Omega} r^7-20 i \ddot{\Omega} \dot{\Omega} r^7\cr&+224 \dot{p}_{r_*} \Omega ^3 r^6-64 \ddot{r} \Omega ^3 r^6+112 i \ddot{p}_{r_*} \Omega ^2 r^6-32 i r^{(3)} \Omega ^2 r^6+28 i p^{(4)}_{r_*} r^6+2 i r^{(5)} r^6+56 p^{(3)}_{r_*} \Omega  r^6-56 \dot{p}_{r_*} \ddot{\Omega} r^6\cr&+16 \ddot{r} \ddot{\Omega} r^6-56 \ddot{p}_{r_*} \dot{\Omega} r^6+16 r^{(3)} \dot{\Omega} r^6+336 i \dot{p}_{r_*} \Omega  \dot{\Omega} r^6-96 i \ddot{r} \Omega  \dot{\Omega} r^6-112 p_\varphi \Omega ^4 r^5+256 \dot{r}^2 \Omega ^3 r^5+224 p_{r_*} \dot{r} \Omega ^3 r^5\cr&+112 i p_{r_*} \ddot{r} \Omega ^2 r^5+896 i \dot{p}_{r_*} \dot{r} \Omega ^2 r^5+192 i \ddot{r} \dot{r} \Omega ^2 r^5+84 p_\varphi \dot{\Omega}^2 r^5+112 i \ddot{p}_{r_*} \ddot{r} r^5+56 i \dot{p}_{r_*} r^{(3)} r^5+12 i \ddot{r} r^{(3)} r^5\cr&+28 i p_{r_*} r^{(4)} r^5+168 i p^{(3)}_{r_*} \dot{r} r^5-10 i r^{(4)} \dot{r} r^5-24 \ddot{r}^2 \Omega  r^5-168 \dot{p}_{r_*} \ddot{r} \Omega  r^5+56 p_{r_*} r^{(3)} \Omega  r^5-56 \ddot{p}_{r_*} \dot{r} \Omega  r^5\cr&-96 r^{(3)} \dot{r} \Omega  r^5-64 \dot{r}^2 \ddot{\Omega} r^5-56 p_{r_*} \dot{r} \ddot{\Omega} r^5+112 p_\varphi \Omega  \ddot{\Omega} r^5+14 i p_\varphi \Omega^{(3)} r^5-336 i p_\varphi \Omega ^2 \dot{\Omega} r^5-56 p_{r_*} \ddot{r} \dot{\Omega} r^5\cr&-448 \dot{p}_{r_*} \dot{r} \dot{\Omega} r^5-96 \ddot{r} \dot{r} \dot{\Omega} r^5+384 i \dot{r}^2 \Omega  \dot{\Omega} r^5+336 i p_{r_*} \dot{r} \Omega  \dot{\Omega} r^5-448 i p_\varphi \dot{r} \Omega ^3 r^4-56 i p_{r_*} \ddot{r}^2 r^4+112 i \ddot{p}_{r_*} \dot{r}^2 r^4\cr&-88 i r^{(3)} \dot{r}^2 r^4+288 i \dot{r}^3 \Omega ^2 r^4+672 i p_{r_*} \dot{r}^2 \Omega ^2 r^4+336 p_\varphi \ddot{r} \Omega ^2 r^4-14 p_\varphi r^{(4)} r^4-84 i \ddot{r}^2 \dot{r} r^4-112 i \dot{p}_{r_*} \ddot{r} \dot{r} r^4\cr&-784 \dot{p}_{r_*} \dot{r}^2 \Omega  r^4-336 \ddot{r} \dot{r}^2 \Omega  r^4+112 i p_\varphi r^{(3)} \Omega  r^4-560 p_{r_*} \ddot{r} \dot{r} \Omega  r^4+112 i p_\varphi \dot{r} \ddot{\Omega} r^4-144 \dot{r}^3 \dot{\Omega} r^4-336 p_{r_*} \dot{r}^2 \dot{\Omega} r^4\cr&+168 i p_\varphi \ddot{r} \dot{\Omega} r^4+672 p_\varphi \dot{r} \Omega  \dot{\Omega} r^4-112 i \dot{p}_{r_*} \dot{r}^3 r^3-24 i \ddot{r} \dot{r}^3 r^3-42 p_\varphi \ddot{r}^2 r^3-224 i p_{r_*} \ddot{r} \dot{r}^2 r^3+336 p_\varphi \dot{r}^2 \Omega ^2 r^3\cr&-288 i p_{r_*} \Omega ^2 r^3+72 i \ddot{p}_{r_*} r^3-56 p_\varphi r^{(3)} \dot{r} r^3-24 \dot{r}^4 \Omega  r^3-336 p_{r_*} \dot{r}^3 \Omega  r^3+288 \dot{p}_{r_*} \Omega  r^3+336 i p_\varphi \ddot{r} \dot{r} \Omega  r^3\cr&+168 i p_\varphi \dot{r}^2 \dot{\Omega} r^3+144 p_{r_*} \dot{\Omega} r^3-576 p_\varphi \Omega ^2 r^2+144 \ddot{p}_\varphi r^2-576 i \dot{p}_\varphi \Omega  r^2-288 i p_\varphi \dot{\Omega} r^2\cr&-144 p_\varphi \ddot{r} r-288 \dot{p}_\varphi \dot{r} r+576 i p_\varphi \dot{r} \Omega  r+288 p_\varphi \dot{r}^2 \bigg),
\end{align}
\end{widetext}
where we stress that $h_{22}^{\rm 2.5PN_{\rm inst}}$ is one of the novel waveform contributions presented in this work.
The instantaneous terms for the other $m\neq0$ spherical modes up to $\l=6$ are collected in the supplementary file, together with the respective expressions that follow when the usual order-reduction of the time derivatives is instead performed.

We specify that the approach we described and adopted above is only suitable for the instantaneous part of the waveform. In fact, to evaluate the time integrals over the past history of the source, which appear in all the hereditary contributions, one needs to make use of the PN-expanded EOM; the same happens for the derivation of the post-adiabatic terms.

\subsection{Quasi-Keplerian harmonic parametrization in EOB coordinates: post-adiabatic and hereditary terms}
\label{sec:EOM_terms}
Throughout this section we will consider the useful spherical mode notation
\begin{equation}
	\label{basicStruct}
	h_\lm=-\frac{8 \nu}{c^4}\sqrt{\dfrac{\pi}{5}} e^{-im\varphi} \hat{H}_{\lm},
\end{equation}
already employed (modulo different numerical factors) in Refs.~\cite{Khalil:2021txt,Placidi:2021rkh}, where $\hat{H}_{\lm}$ has been computed in EOB coordinates up to the 2PN order, without including memory terms.
In this section we go through the  derivation of the EOB-coordinate expressions for $\hat{H}_{\lm}$ up to the 2.5PN order while including all the waveform contributions that are inherently left out from the procedure outlined in the previous section: hereditary (tail + memory) and post-adiabatic terms. We stress in this respect that the memory terms we are considering here specifically belong to the so called \emph{oscillatory memory}, a component of the nonlinear memory \cite{Christodoulou:1991cr,Wiseman:1991ss,Blanchet:1992br} whose hereditary time integrals involve oscillatory exponentials. Since these oscillations tend to cancel each other out going towards the remote past of the dynamical history of the source, these contributions are more similar to the tail than to the genuine non-oscillatory memory, the so-called \emph{direct current} memory, present only for the $m=0$ spherical modes whose analysis we leave for future work.

In general, the waveform contributions we are interested in here were derived up to the 3PN order in Ref.~\cite{Ebersold:2019kdc}, using the quasi-Keplerian (QK) parametrization \cite{Damour:1985,Memmesheimer:2004cv} and performing an expansion for small time eccentricity $e_t$ up to order $O(e_t^6)$. Our aim is to recast these results in EOB coordinates, so to get them ready for the implementation into EOB models, limiting ourselves to 2.5PN accuracy. To this end, starting from the QK expressions of Ref.~\cite{Ebersold:2019kdc}, we need to (i) recover the usual harmonic coordinates from the QK orbital parameters and (ii) use the proper 2PN accurate\footnote{The next PN order after the 2PN in the transformations from harmonic to EOB coordinates is the 3PN. Therefore, for a 2.5PN accurate waveform, the 2PN accurate coordinate transformations are all we need.} coordinate transformations to finally recast everything in EOB coordinates.
Before we start, it should be noted that all the contributions we consider in this section enter the complete waveform starting from the 1.5PN order. This brings a huge simplification in the procedure outlined above, especially in its first step.

In the following we outline in detail the entire translation procedure while applying it step-wise, as an illustrative example, to the leading oscillatory memory term in the PN expansion of the $\ell=m=2$ spherical mode.  The latter is provided in its original QK form, together with the other contributions we want to translate in EOB coordinates, in the Supplementary Material of Ref.~\cite{Ebersold:2019kdc}; up to order $O(e_t^6)$ it reads
\begin{widetext}
	\begin{align}\label{H22memind}
		& \hat{H}^{\rm 1.5PN_{\rm mem}}_{22}=x^{3/2} i \nu \left[\frac{13  e^{-2 i \xi
		}}{6048}e_t^6+\frac{1495  e^{6 i \xi
		}}{4032}e_t^6 +\frac{55}{504} e_t^6  +\frac{13 
			e^{-i \xi }}{3024} e_t^5+\frac{767   e^{5 i
				\xi }}{3024} e_t^5+\frac{13}{336}  e_t^4  \right.
		\cr
		&\left.-e^{4 i \xi }
		\left(\frac{169   }{1512}e_t^6+\frac{169  
		}{1008}e_t^4\right)-e^{3 i \xi } \left(\frac{25}{336}
		e_t^5 +\frac{13}{126} e_t^3 \right)-e^{i \xi
		} \left(\frac{185 }{1008} e_t^5-\frac{13}{126} 
		e_t^3  \right)\right.
		\cr
		&\left.-e^{2 i \xi } \left(\frac{503  
		}{4032}e_t^6-\frac{5}{72}  e_t^4 +\frac{13}{252} 
		e_t^2  \right)\right].
	\end{align}
\end{widetext}
Here $x=\left(G M\omega/c^3\right)^{2/3}$ is a PN-counting frequency parameter 
defined in terms of the orbital frequency $\omega=(1+k)n$, where $1+k=2\pi/\Phi$ gives the angle advance of the periastron per revolution and $n=2\pi/P$ is the mean motion associated with the period $P$; 
finally the phase angle $\xi$ is a redefinition of the mean anomaly $l$ which arises from a shift of the time coordinate\footnote{This same shift also results in a redefinition of the harmonic phase but it enters the latter as a 4PN correction; we can thus ignore it for our 2.5PN-accurate computation.} aimed at removing from the instantaneous and tail parts of the waveform the arbitrary parameter $x_0$ introduced in the multipolar waveform generation formalism (see, \emph{e.g.}, Ref,~\cite{Blanchet:2013haa}). At 1PN order one simply has
\begin{equation}
	\label{xi}
	\xi= l = u_e -e_t \sin u_e~.
\end{equation}
The parameter $u_e$ above is the eccentric anomaly which enters the Keplerian parametrization of elliptic orbits in polar coordinates and in the center of mass frame, that is
\begin{subequations}
	\label{eccentricmotion}
	\begin{align}
		\label{QPr}
		&  R_h=a(1-e \cos u_e)~,\\
		\label{QPphi}
		&  \varphi_h-(\varphi_{h})_0= 2\arctan\left[\left(\frac{1+e}{1-e}\right)^{1/2}\tan\frac{u_e}{2}\right]~,
	\end{align}
\end{subequations}
where also the semi-major axis $a$ appears, while $e$ is the Newtonian orbital eccentricity; $R_h$ and $\varphi$ define the components of the relative separation $\vec{R_h}=R_h(\cos\varphi_h,\sin\varphi_h,0)$ and the subscript ``$h$" signals that we are employing harmonic coordinates. The QK parametrization used in Ref.~\cite{Ebersold:2019kdc} is non other than a post-Newtonian generalization of the Keplerian parametrization outlined above, and reduces to it when truncated at the Newtonian order. At 1PN, the profile of the parametrization remains the same as in Eq.~\eqref{eccentricmotion} but one has two other eccentricities in addition to $e_t$ appearing in Eq.~\eqref{xi}: the radial eccentricity $e_r$ and the angular eccentricity $e_\varphi$, respectively replacing the Newtonian eccentricity $e$ in Eq.~\eqref{QPr} and \eqref{QPphi}.\footnote{At the leading Newtonian order all the various types of eccentricities coincide and reduce to $e$.} 

Starting from the 2PN order one also has additional terms appearing in Eq.~\eqref{QPphi}. However since the accuracy we work with is 2.5PN and the expressions we need to translate in EOB coordinates enter at 1.5PN order,  we just need the coordinate transformation between the QK parametrization and the harmonic coordinates at 1PN order.
Therefore we can safely use Eqs.~\eqref{xi} and the 1PN-corrected version of  \eqref{eccentricmotion}, identifying $R_h$ and $\varphi_h$ as the harmonic radial separation and phase,
together with the 1PN relations \cite{Memmesheimer:2004cv}
\begin{subequations}\label{orbitalelements}
	\begin{align}
		&a=-\frac{1}{2 E}\bigg[1-\frac{ E}{2}\frac{1}{c^2} (-7+\nu)\bigg],\\
		&e_t^2=1+2 E J^2-\frac{1}{c^2}\frac{E}{2}\bigg[8(-1+\nu)+2 E J^2 (-17+7 \nu)\bigg],\\
		&e_r^2=1+2 E J^2-\frac{1}{c^2}\frac{E}{2}\bigg[ 4(6-\nu)-10 E J^2 (-3+\nu ) \bigg],\\
		&n=(-2 E)^{3/2} \bigg[1+\frac{1}{c^2}\bigg(\frac{ E}{4}(15-\nu)\bigg)\bigg],\\
		&x=-\frac{2E}{c^2}\bigg[1-\frac{2}{3 c^2} \bigg(\frac{3}{J^2}-\frac{1}{4} E(\nu -15)\bigg)\bigg],
	\end{align}
\end{subequations}
which connect the orbital elements we need to rewrite in harmonic coordinates with the binary orbital energy and angular momentum per unit reduced mass $\mu$, respectively denoted as $E$ and $J$. 
Then, using the mass reduced radial coordinate $r_h \equiv R_h/M$, at 1PN we have
\begin{align}
	\label{EandJ}
	E&= \frac{\dot r_h^2}{2}+\frac{1}{2} r_h^2\Omega_h^2-\frac{1}{r_h}+\frac{1}{c^2}\bigg[ \frac{3}{8}(1-3 \nu) \big(\dot{r_h}^2+r_h^2 \dot{\varphi_h}^2\big)^2\cr 
	&+\frac{1}{2} \nu \frac{\dot{r_h}^2}{r_h} + \frac{(3+\nu)(\dot{r_h}^2+r_h^2 \dot{\varphi_h}^2)}{2r_h}+\frac{1}{2r_h^2}\bigg],\\
	J &=r_h^2 \dot{\varphi_h} +\frac{r_h^2 \dot{\varphi_h}}{c^2} \bigg[\frac{1}{2}(1-3 \nu) (\dot{r_h}^2+r_h^2 \dot{\varphi_h}^2)\cr
	&+ \frac{3+\nu}{r_h}\bigg].
\end{align}

By combining Eqs.~\eqref{xi}-\eqref{EandJ} we can thus rewrite as desired the orbital elements $u_e$, $e$ and $x$ in terms of the harmonic polar coordinates $r_h$, $\varphi_h$ and their first time derivatives $\dot r$ and $\Omega_h \equiv \dot \varphi_h$.
The 1PN explicit expression we find are
\begin{widetext}
\begin{align}
	u_e
	&=\arccos{\Bigg[\frac{-1+r_h \dot{r}_h^2+r_h^3 \Omega_h^2}{\sqrt{1-2 r_h^3 \Omega_h^2+r_h^4 \dot{r}_h^2 \Omega_h^2+r_h^6 \Omega_h^4}}\Bigg]}- \frac{1}{c^2}\frac{\dot{r}_h^2}{2}\frac{2+r_h \dot{r}_h^2 (4-5 \nu)+5 \nu+r_h^3 (4-8 \nu+r_h \dot{r}_h^2 \nu) \Omega_h^2 +r_h^6 \nu \Omega_h^4}{ \sqrt{r_h \dot{r}_h^2 (2-r_h \dot{r}_h^2 -r_h^3 \Omega_h^2)} \big(1-2 r_h^3 \Omega_h^2+r_h^4 \dot{r}_h^2 \Omega_h^2+r_h^6 \Omega_h^4\big)},\\
	e_t&=\sqrt{1-2 r_h^3 \Omega_h^2+r_h^4 \dot{r}_h^2 \Omega_h^2 +r_h^6 \Omega_h^4} +\frac{1}{c^2}\frac{1}{2 r_h \sqrt{1-2 r_h^3 \Omega_h^2+r_h^4 \dot{r}_h^2 \Omega_h^2 +r_h^6 \Omega_h^4}} \Bigg[2 r_h \dot{r}_h^2+4 (-1+\nu) -2 r_h \dot{r}_h^2 \nu\cr 
	&+r_h^3 (8-13 \nu) \Omega_h^2  +r_h^5 \dot{r}_h^4 (6-7 \nu) \Omega_h^2+ r_h^4 \dot{r}_h^2 (-10+17 \nu) \Omega_h^2+2 r_h^7 \dot{r}_h^2 (6-7 \nu) \Omega_h^4+2 r_h^6 (-5+8 \nu) \Omega_h^4\cr
	&+r_h^9 (6-7 \nu) \Omega_h^6\Bigg],\\
	x &=\frac{2-r_h \dot{r}_h^2-r_h^3 \Omega_h^2}{r_h}+ \frac{1}{c^2} \frac{1}{3 r_h^5 \Omega_h^2} \Bigg[12-6 r_h \dot{r}_h^2+
	r_h^4 \dot{r}_h^2 (6-7 \nu) \Omega_h^2+
	r_h^3 (-24+\nu) \Omega_h^2 +
	r_h^5 \dot{r}_h^4 (-6+7 \nu) \Omega_h^2\cr
	&+
	2 r_h^6 (3-2 \nu) \Omega_h^4+
	2 r_h^7 \dot{r}_h^2 (-6+7 \nu) \Omega_h^4+
	r_h^9 (-6+7 \nu) \Omega_h^6\Bigg].
\end{align}
\end{widetext}

An important consideration is in order: in the leading Keplerian motion, one has 
\begin{subequations}\label{rdotphidot}
	\begin{align}
		&\dot r_h=  e\, \frac{n^{1/3}\sin u_e}{1-e \cos u_e}~,\\
		&\Omega_h= n \frac{\sqrt{1-e^2}}{(1-e\cos u_e)^2}~,
	\end{align}
\end{subequations}
This shows as expected that the quantity $\dot{r}_h$ is proportional to the eccentricity $e$ and thus goes to zero in the circular limit.
In addition to this, it is possible to construct yet another quantity proportional to the eccentricity: introducing the variable $k_p \equiv 1-\Omega_h^2 r_h^3$ and using  Eq.~\eqref{rdotphidot} yields
\begin{equation}\label{kp}
	k_p =e\frac{e-\cos u_e}{1-e \cos u_e},
\end{equation}
in compliance with the fact that $\Omega_h^2 r_h^3=1$ when the motion is circular.
Therefore, to properly keep into account the eccentricity expansion of Eq.~\eqref{H22memind} in our translation into harmonic coordinates, we must rescale $\dot r_h$ and $k_p$ by a small parameter $\epsilon$, which we use to properly keep track of the power order in eccentricity, and expand in $\epsilon$ up to $\mathcal{O}(\epsilon^6)$. This produces a simultaneous expansion in $\dot r_h$ and $k_p$ up to the sixth order, the same order at which the expansion in eccentricity underlying the starting QK contributions to the waveform multipoles, $e.g.$ Eq.\eqref{H22memind}, is truncated. In this expansion one would get expressions containing also half integer powers of $r_h$, although they can be removed by harnessing the definition of the quantity $k_p$. We can in fact use the latter to rewrite $\sqrt{r_h}$ as
\begin{align}
	\sqrt{r_h}&=\frac{r_h^2 \Omega_h }{\sqrt{1-k_p \epsilon }}=r_h^2 \Omega_h\bigg(1+\frac{k_p \epsilon }{2}+\frac{3 k_p^2 \epsilon ^2}{8}+\frac{5 k_p^3 \epsilon ^3}{16}\cr
	&+\frac{35 k_p^4 \epsilon ^4}{128}+\frac{63 k_p^5 \epsilon ^5}{256}+\frac{231 k_p^6 \epsilon ^6}{1024}\bigg)+O\big(\epsilon ^7\big),
\end{align}
and use it in our expansion in $\epsilon$. After this step we set $\epsilon$ to 1 and rewrite $k_p$ in terms of $r_h$ and $\Omega_h$,
so that we finally end up with expressions that depend only on the harmonic coordinates $r_h$, $\dot{r}_h$ and $\Omega_h$. Coming back to Eq.~\eqref{H22memind}, with the procedure described above it becomes
\begin{widetext}
	\begin{align}
		&\hat{H}^{\rm 1.5PN_{\rm mem}}_{22}=\frac{\nu}{c^3}  \Bigg(\frac{23}{28}  i r_h^{17} \Omega_h^{13}-\frac{689}{126} i r_h^{14} \Omega_h^{11}-\frac{71}{126} r_h^{13} \dot{r}_h \Omega_h^{10}-\frac{65}{252} i r_h^{12} \dot{r}_h^2 \Omega_h^9+\frac{3865}{252} i r_h^{11} \Omega_h^9+\frac{137}{42} r_h^{10} \dot{r}_h \Omega_h^8\cr
		&+\frac{26}{21} i r_h^9 \dot{r}_h^2 \Omega_h^7-\frac{1}{63} r_h^8 \dot{r}_h^3 \Omega_h^6-\frac{93}{4} i r_h^8 \Omega_h^7-\frac{1}{126} i r_h^7 \dot{r}_h^4 \Omega_h^5-\frac{325}{42} r_h^7 \dot{r}_h \Omega_h^6-\frac{65}{28} i r_h^6 \dot{r}_h^2 \Omega_h^5+\frac{4}{63} r_h^5 \dot{r}_h^3 \Omega_h^4+\frac{725}{36} i r_h^5 \Omega_h^5\cr
		&+\frac{1}{42} i r_h^4 \dot{r}_h^4 \Omega_h^3+\frac{1195}{126} r_h^4 \dot{r}_h \Omega_h^4+\frac{130}{63} i r_h^3 \dot{r}_h^2 \Omega_h^3-\frac{2}{21} r_h^2 \dot{r}_h^3 \Omega_h^2+\frac{23 \dot{r}_h}{14 r_h^2}-\frac{265}{28} i r_h^2 \Omega_h^3-\frac{1}{42} i r_h \dot{r}_h^4 \Omega_h+\frac{\dot{r}_h^3}{21 r_h}-\frac{767}{126} r_h \dot{r}_h \Omega_h^2+\cr
		&+\frac{475 i \Omega_h}{252 r_h}-\frac{65}{84} i \dot{r}_h^2 \Omega_h\Bigg).
	\end{align}
\end{widetext}
We can now trade the harmonic coordinates $(r_h,~\dot r_h,~\varphi_h,~\Omega_h)$
with the mass-reduced EOB phase-space variables $(r, \varphi,p_{r_*},p_\varphi)$ at 2PN accuracy, by employing the transformations shown in Eqs.~(5)-(8) of Ref.~\cite{Placidi:2021rkh}.  The final result for the leading oscillatory memory terms we are explicitly computing as an illustrative example of the procedure is
\begin{widetext}
\begin{align}
	\label{eob_leading_osc_mem}
	\hat{H}^{\rm 1.5PN_{\rm mem}}_{22} &= \frac{\nu}{c^3}\Bigg(-\frac{71}{126} u^7 p_{r_*} p_{\varphi }^{10}-\frac{65}{252} i u^6 p_{r_*}^2 p_{\varphi }^9+\frac{137}{42} u^6 p_{r_*} p_{\varphi }^8+\frac{26}{21} i u^5 p_{r_*}^2 p_{\varphi }^7-\frac{325}{42} u^5 p_{r_*} p_{\varphi }^6-\frac{1}{63} u^4 p_{r_*}^3 p_{\varphi }^6\cr
	&-\frac{65}{28} i u^4 p_{r_*}^2 p_{\varphi }^5+\frac{1195}{126} u^4 p_{r_*} p_{\varphi }^4-\frac{1}{126} i u^3 p_{r_*}^4 p_{\varphi }^5+\frac{4}{63} u^3 p_{r_*}^3 p_{\varphi }^4+\frac{130}{63} i u^3 p_{r_*}^2 p_{\varphi }^3-\frac{767}{126} u^3 p_{r_*} p_{\varphi }^2+\frac{1}{42} i u^2 p_{r_*}^4 p_{\varphi }^3\cr
	&-\frac{2}{21} u^2 p_{r_*}^3 p_{\varphi }^2-\frac{65}{84} i u^2 p_{r_*}^2 p_{\varphi }+\frac{23}{14} u^2 p_{r_*}-\frac{1}{42} i u p_{r_*}^4 p_{\varphi }+\frac{1}{21} u p_{r_*}^3+\frac{23}{28} i u^9 p_{\varphi }^{13}-\frac{689}{126} i u^8 p_{\varphi }^{11}+\frac{3865}{252} i u^7 p_{\varphi }^9\cr
	&-\frac{93}{4} i u^6 p_{\varphi }^7+\frac{725}{36} i u^5 p_{\varphi }^5-\frac{265}{28} i u^4 p_{\varphi }^3+\frac{475}{252} i u^3 p_{\varphi }\Bigg),
\end{align}
\end{widetext}
where we introduced the variable $u\equiv 1/r$. The final expressions in EOB coordinates for all the hereditary and post-adiabatic contributions to the waveform up to 2.5PN order can be found in the Mathematica notebook that accompanies this paper, encompassing each relevant spherical mode with $m\neq0$. Below, focusing on the mode $\ell=m=2$, we limit ourselves to reporting the EOB-coordinate contributions that represent a novelty in the EOB literature, in addition to the leading order oscillatory memory already given in Eq.~\eqref{eob_leading_osc_mem}. These are the 2.5PN oscillatory memory,
\begin{widetext}
\begin{align}
    \hat{H}^{\rm 2.5PN_{\rm mem}}_{22}&= \frac{\nu}{c^5} \bigg[\left(\frac{345 \nu }{28}-\frac{299}{56}\right) i u^{11} p_{\varphi}^{15}+\frac{23}{14} p_{r_*} u^{10} \nu  p_{\varphi}^{14}+\left(\frac{845099 \nu }{8064}-\frac{5339291}{48384}\right) i u^{10} p_{\varphi}^{13}\cr&+\left(\frac{69 \nu }{14}-\frac{299}{56}\right) i p_{r_*}^2 u^9 p_{\varphi}^{13}+\left(\frac{781}{252}-\frac{4673 \nu }{252}\right) p_{r_*} u^9 p_{\varphi}^{12}+\left(\frac{19990273}{24192}-\frac{3973649 \nu }{4032}\right) i u^9 p_{\varphi}^{11} \cr& +\left(\frac{1598621}{43008}-\frac{797269 \nu }{21504}\right) i p_{r_*}^2 u^8 p_{\varphi}^{11}+\left(\frac{100321}{896}-\frac{45331 \nu }{448}\right) p_{r_*} u^8 p_{\varphi}^{10}+\left(\frac{781}{252}-\frac{911 \nu }{252}\right) p_{r_*}^3 u^7 p_{\varphi}^{10}\cr&+\left(\frac{3421043 \nu }{1152}-\frac{37301543}{16128}\right) i u^8 p_{\varphi}^9+\left(\frac{2526931 \nu }{64512}-\frac{4983163}{129024}\right) i p_{r_*}^2 u^7 p_{\varphi}^9+\left(\frac{715}{504}-\frac{845 \nu }{504}\right) i p_{r_*}^4 u^6 p_{\varphi}^9\cr&+\left(\frac{3354823 \nu }{4032}-\frac{1784815}{2688}\right) p_{r_*} u^7 p_{\varphi}^8+\left(\frac{965 \nu }{63}-\frac{409}{28}\right) p_{r_*}^3 u^6 p_{\varphi}^8+\left(\frac{41009279}{12096}-\frac{1023047 \nu }{224}\right) i u^7 p_{\varphi}^7\cr&+\left(\frac{5064533 \nu }{32256}-\frac{8498621}{64512}\right) i p_{r_*}^2 u^6 p_{\varphi}^7+\left(-\frac{2515 \nu }{4032}-\frac{6085}{4032}\right) i p_{r_*}^4 u^5 p_{\varphi}^7+\left(\frac{2600881}{1728}-\frac{4153495 \nu }{2016}\right) p_{r_*} u^6 p_{\varphi}^6\cr&+\left(\frac{142937}{4032}-\frac{92747 \nu }{2016}\right) p_{r_*}^3 u^5 p_{\varphi}^6+\left(\frac{1}{14}-\frac{\nu }{9}\right) p_{r_*}^5 u^4 p_{\varphi}^6+\left(\frac{31548677 \nu }{8064}-\frac{135064421}{48384}\right) i u^6 p_{\varphi}^5\cr&+\left(\frac{7486247}{21504}-\frac{222275 \nu }{512}\right) i p_{r_*}^2 u^5 p_{\varphi}^5+\left(\frac{282325}{16128}-\frac{11251 \nu }{2688}\right) i p_{r_*}^4 u^4 p_{\varphi}^5+\left(\frac{1}{28}-\frac{\nu }{18}\right) i p_{r_*}^6 u^3 p_{\varphi}^5\cr&+\left(\frac{1659109 \nu }{672}-\frac{6897269}{4032}\right) p_{r_*} u^5 p_{\varphi}^4+\left(\frac{65339 \nu }{672}-\frac{717665}{12096}\right) p_{r_*}^3 u^4 p_{\varphi}^4+\left(\frac{\nu }{3}-\frac{2}{9}\right) p_{r_*}^5 u^3 p_{\varphi}^4\cr&+\left(\frac{9928091}{8064}-\frac{7213553 \nu }{4032}\right) i u^5 p_{\varphi}^3+\left(\frac{26373029 \nu }{64512}-\frac{39590701}{129024}\right) i p_{r_*}^2 u^4 p_{\varphi}^3+\left(\frac{5473 \nu }{224}-\frac{990629}{24192}\right) i p_{r_*}^4 u^3 p_{\varphi}^3\cr&+\left(\frac{1369}{16128}-\frac{1921 \nu }{24192}\right) i p_{r_*}^6 u^2 p_{\varphi}^3+\left(\frac{2655167}{2688}-\frac{5998967 \nu }{4032}\right) p_{r_*} u^4 p_{\varphi}^2+\left(\frac{103847}{1728}-\frac{71987 \nu }{672}\right) p_{r_*}^3 u^3 p_{\varphi}^2\cr&+\left(\frac{5077 \nu }{4032}-\frac{44519}{8064}\right) p_{r_*}^5 u^2 p_{\varphi}^2+\left(\frac{2763659 \nu }{8064}-\frac{3673177}{16128}\right) i u^4 p_{\varphi}+\left(\frac{4278091}{43008}-\frac{1024033 \nu }{7168}\right) i p_{r_*}^2 u^3 p_{\varphi}\cr&+\left(\frac{417677}{16128}-\frac{174901 \nu }{8064}\right) i p_{r_*}^4 u^2 p_{\varphi}+\left(\frac{97}{72}-\frac{773 \nu }{1512}\right) i p_{r_*}^6 u p_{\varphi}+\left(\frac{1472603 \nu }{4032}-\frac{5644187}{24192}\right) p_{r_*} u^3\cr&+\left(\frac{89099 \nu }{2016}-\frac{99881}{4032}\right) p_{r_*}^3 u^2+\left(\frac{45991}{8064}-\frac{1991 \nu }{1344}\right) p_{r_*}^5 u \bigg],
\end{align}
the 2.5PN tail
\begin{align}
    \hat{H}^{\rm 2.5PN_{\rm tail}}_{22}&=\frac{1}{c^5}\bigg\{ \frac{3}{2} i p_{r_*}^6 p_{\varphi}^3 u^2+\frac{9 p_{r_*}^5 u}{2}-\frac{405}{16} i p_{r_*}^4 p_{\varphi}^7 u^5+\frac{783}{16} i p_{r_*}^4 p_{\varphi}^5 u^4-\frac{27}{16} i p_{r_*}^4 p_{\varphi}^3 u^3-\frac{567}{16} i p_{r_*}^4 p_{\varphi} u^2+9 p_{r_*}^3 p_{\varphi}^6 u^5\cr&-36 p_{r_*}^3 p_{\varphi}^4 u^4+63 p_{r_*}^3 p_{\varphi}^2 u^3-54 p_{r_*}^3 u^2+\frac{945}{16} i p_{r_*}^2 p_{\varphi}^{11} u^8-\frac{3573}{16} i p_{r_*}^2 p_{\varphi}^9 u^7+\frac{1905}{8} i p_{r_*}^2 p_{\varphi}^7 u^6+\frac{711}{8} i p_{r_*}^2 p_{\varphi}^5 u^5\cr&-\frac{5427}{16} i p_{r_*}^2 p_{\varphi}^3 u^4+\frac{3207}{16} i p_{r_*}^2 p_{\varphi} u^3-6 p_{r_*} p_{\varphi}^{10} u^8+\frac{75}{2} p_{r_*} p_{\varphi}^8 u^7-99 p_{r_*} p_{\varphi}^6 u^6+147 p_{r_*} p_{\varphi}^4 u^5-138 p_{r_*} p_{\varphi}^2 u^4\cr&+\frac{135 p_{r_*} u^3}{2}-39 i p_{\varphi}^{13} u^{10}+\frac{1077}{4} i p_{\varphi}^{11} u^9-\frac{3171}{4} i p_{\varphi}^9 u^8+\frac{2571}{2} i p_{\varphi}^7 u^7-\frac{2463}{2} i p_{\varphi}^5 u^6+\frac{2733}{4} i p_{\varphi}^3 u^5-\frac{699}{4} i p_{\varphi} u^4\cr&+ \pi \bigg[ \left(\frac{1261}{1152}-\frac{485 \nu }{192}\right) u^{11} p_{\varphi}^{15}+\frac{97}{288} i p_{r_*} u^{10} \nu  p_{\varphi}^{14}+\left(\frac{23951045}{24192}-\frac{3174763 \nu }{10080}\right) u^{10} p_{\varphi}^{13}\cr&+\left(\frac{1261}{1152}-\frac{97 \nu }{96}\right) p_{r_*}^2 u^9 p_{\varphi}^{13}+\left(-\frac{627 \nu }{320}-\frac{11}{64}\right) i p_{r_*} u^9 p_{\varphi}^{12}+\left(\frac{26112923 \nu }{13440}-\frac{30040091}{5040}\right) u^9 p_{\varphi}^{11}\cr&+\left(\frac{2309 \nu }{640}-\frac{3751}{640}\right) p_{r_*}^2 u^8 p_{\varphi}^{11}+\left(\frac{3416687}{20160}-\frac{265669 \nu }{5376}\right) i p_{r_*} u^8 p_{\varphi}^{10}+\left(\frac{27 \nu }{64}-\frac{11}{64}\right) i p_{r_*}^3 u^7 p_{\varphi}^{10}\cr&+\left(\frac{2504909}{168}-\frac{4392159 \nu }{896}\right) u^8 p_{\varphi}^9+\left(\frac{62367 \nu }{896}-\frac{152531}{672}\right) p_{r_*}^2 u^7 p_{\varphi}^9+\left(\frac{11}{16}-\frac{13 \nu }{16}\right) p_{r_*}^4 u^6 p_{\varphi}^9\cr&+\left(\frac{2136049 \nu }{8064}-\frac{3428447}{4032}\right) i p_{r_*} u^7 p_{\varphi}^8+\left(\frac{457}{256}-\frac{2735 \nu }{576}\right) i p_{r_*}^3 u^6 p_{\varphi}^8+\left(\frac{13237351 \nu }{2016}-\frac{480792295}{24192}\right) u^7 p_{\varphi}^7\cr&+\left(\frac{7719}{8}-\frac{14143 \nu }{48}\right) p_{r_*}^2 u^6 p_{\varphi}^7+\left(\frac{1585 \nu }{288}-\frac{435}{128}\right) p_{r_*}^4 u^5 p_{\varphi}^7+\left(\frac{191879}{112}-\frac{1446995 \nu }{2688}\right) i p_{r_*} u^6 p_{\varphi}^6\cr&+\left(\frac{2158799}{12096}-\frac{24457 \nu }{504}\right) i p_{r_*}^3 u^5 p_{\varphi}^6+\left(\frac{137}{128}-\frac{23 \nu }{12}\right) i p_{r_*}^5 u^4 p_{\varphi}^6+\left(\frac{120068783}{8064}-\frac{6637373 \nu }{1344}\right) u^6 p_{\varphi}^5\cr&+\left(\frac{49939 \nu }{112}-\frac{1345559}{896}\right) p_{r_*}^2 u^5 p_{\varphi}^5+\left(\frac{38155 \nu }{1344}-\frac{300511}{2688}\right) p_{r_*}^4 u^4 p_{\varphi}^5+\left(\frac{329 \nu }{192}-\frac{141}{128}\right) p_{r_*}^6 u^3 p_{\varphi}^5\cr&+\left(\frac{728641 \nu }{1344}-\frac{6961777}{4032}\right) i p_{r_*} u^5 p_{\varphi}^4+\left(\frac{152197 \nu }{896}-\frac{4288559}{8064}\right) i p_{r_*}^3 u^4 p_{\varphi}^4+\left(\frac{2519 \nu }{384}-\frac{2331}{640}\right) i p_{r_*}^5 u^3 p_{\varphi}^4\cr&+\left(\frac{26566559 \nu }{13440}-\frac{23966723}{4032}\right) u^5 p_{\varphi}^3+\left(\frac{2785247}{2688}-\frac{267235 \nu }{896}\right) p_{r_*}^2 u^4 p_{\varphi}^3+\left(\frac{901757}{4032}-\frac{62701 \nu }{896}\right) p_{r_*}^4 u^3 p_{\varphi}^3\cr&+\left(\frac{301}{128}-\frac{7139 \nu }{1920}\right) p_{r_*}^6 u^2 p_{\varphi}^3+\left(\frac{3473249}{4032}-\frac{21669611 \nu }{80640}\right) i p_{r_*} u^4 p_{\varphi}^2+\left(\frac{513145}{1008}-\frac{356717 \nu }{2016}\right) i p_{r_*}^3 u^3 p_{\varphi}^2\cr&+\left(-\frac{112363 \nu }{26880}-\frac{95339}{13440}\right) i p_{r_*}^5 u^2 p_{\varphi}^2+\left(\frac{977 \nu }{1920}-\frac{161}{640}\right) i p_{r_*}^7 u p_{\varphi}^2+\left(\frac{7}{96}-\frac{5 \nu }{32}\right) p_{r_*}^8 p_{\varphi}\cr&+\left(\frac{8603621}{8640}-\frac{1906957 \nu }{5760}\right) u^4 p_{\varphi}+\left(\frac{47971 \nu }{640}-\frac{187417}{720}\right) p_{r_*}^2 u^3 p_{\varphi}+\left(\frac{296903 \nu }{8064}-\frac{47083}{448}\right) p_{r_*}^4 u^2 p_{\varphi}\cr&+\left(\frac{14869 \nu }{4480}-\frac{94477}{20160}\right) p_{r_*}^6 u p_{\varphi}+\left(\frac{227}{768}-\frac{227 \nu }{384}\right) i p_{r_*}^7+\left(\frac{150365 \nu }{2688}-\frac{1169549}{6720}\right) i p_{r_*} u^3\cr&+\left(\frac{482947 \nu }{8064}-\frac{7597547}{48384}\right) i p_{r_*}^3 u^2+\left(\frac{46811}{5760}-\frac{841 \nu }{1920}\right) i p_{r_*}^5 u \bigg]\bigg\},
\end{align}
and the leading post-adiabatic contribution, also at 2.5PN,
\begin{align}
\label{eq:post_ad_22}
    \hat{H}^{\rm 2.5PN_{\rm post-ad}}_{22}&= \frac{\nu}{c^5} \bigg(\frac{53811}{80} i p_{r_*}^6 p_{\varphi} u+\frac{4483193}{800} p_{r_*}^5 p_{\varphi}^2 u^2-\frac{12206171 p_{r_*}^5 u}{1800}-\frac{55971121 i p_{r_*}^4 p_{\varphi}^5 u^4}{3600}\cr&+\frac{114780787 i p_{r_*}^4 p_{\varphi}^3 u^3}{3600}-\frac{10008601}{600} i p_{r_*}^4 p_{\varphi} u^2-\frac{6222059}{300} p_{r_*}^3 p_{\varphi}^6 u^5+\frac{214692649 p_{r_*}^3 p_{\varphi}^4 u^4}{3600}\cr&-\frac{34571749}{600} p_{r_*}^3 p_{\varphi}^2 u^3+\frac{21785401 p_{r_*}^3 u^2}{1200}+\frac{28997147 i p_{r_*}^2 p_{\varphi}^9 u^7}{3600}-\frac{55166819 i p_{r_*}^2 p_{\varphi}^7 u^6}{1800}+\frac{39128203}{900} i p_{r_*}^2 p_{\varphi}^5 u^5\cr&-\frac{48531559 i p_{r_*}^2 p_{\varphi}^3 u^4}{1800}+\frac{21851597 i p_{r_*}^2 p_{\varphi} u^3}{3600}+\frac{138427943 p_{r_*} p_{\varphi}^{10} u^8}{36000}-\frac{75947419 p_{r_*} p_{\varphi}^8 u^7}{3600}\cr&+\frac{166293283 p_{r_*} p_{\varphi}^6 u^6}{3600}-\frac{46017217}{900} p_{r_*} p_{\varphi}^4 u^5+\frac{207786223 p_{r_*} p_{\varphi}^2 u^4}{7200}-\frac{13558101 p_{r_*} u^3}{2000}+\frac{14097919 i p_{\varphi}^{13} u^{10}}{18000}\cr&-\frac{89224889 i p_{\varphi}^{11} u^9}{18000}+\frac{6070009}{450} i p_{\varphi}^9 u^8-\frac{4017987}{200} i p_{\varphi}^7 u^7+\frac{62222567 i p_{\varphi}^5 u^6}{3600}-\frac{146841689 i p_{\varphi}^3 u^5}{18000}\cr&+\frac{14952347 i p_{\varphi} u^4}{9000}\bigg).
\end{align}
\end{widetext}

We highlight that all the oscillatory memory terms are purely non-circular contributions\footnote{In fact they stem from QK expressions like Eq.~\eqref{H22memind} whose terms are all proportional to the eccentricity $e$.} that are proportional to the symmetric mass ratio $\nu$. This means that they disappear both in the test-mass limit $\nu\to0$ and in the circular limit, so that the impact of their eventual inclusion in EOB models can be only assessed by performing comparisons with numerically simulated waveform for comparable-mass non-circularized binaries. Regarding the post-adiabatic term \eqref{eq:post_ad_22} we specify instead that, while it is also proportional to $\nu$, it survives in the circular limit and its contribution is actually needed to correctly reproduce the quasi-circular PN expression of $h_{22}$ given, e.g.,  in Ref.~\cite{Kidder:2007rt}.\footnote{In particular, it can be seen that the 2.5PN-accurate term $-24 i \nu  x^{5/2}$ inside the curl brackets in Eq.~(79) of Ref.~\cite{Kidder:2007rt} results form the sum between the circular limits of the instantaneous contributions \eqref{eq:Newt_prefactor}-\eqref{eq:inst2.5PN} and the post-adiabatic term \eqref{eq:post_ad_22}.}

\section{Generic-orbit 2.5PN factorized waveform}
\label{sec:nc_factors}
The spherical multipoles of the waveform derived in EOB coordinates in the previous section at 2.5PN accuracy have the following structure:
\begin{align}
	&h_\lm  = h_\lm^{\rm inst}+h_\lm^{\rm QK},
\end{align}
where (i) $h_\lm^{\rm inst}$ is the purely instantaneous part of the mode, derived in Sec.~\ref{sec:InstNovVague} and presenting a PN profile of the type \eqref{h22_inst}, which also encompasses the leading Newtonian term; the component $h_\lm^{\rm QK}$ addresses instead the set of all the other contributions to the spherical mode, computed from QK waveform results as outlined in Sec.~\ref{sec:EOM_terms}, and it is always subleading with respect to the Newtonian term. For instance, in the case of the dominant $\ell=m=2$ mode, one has\footnote{Here each waveform piece is considered in its full form, obtained by multiplying the corresponding $\hat{H}_\lm$ of the previous section with the prefactor of Eq.~\eqref{basicStruct}.}
\begin{align}
	&h_{22}^{\rm QK} = \frac{1}{c^3}\left( h_{22}^{\rm 1.5PN_{\rm tail}}+h_{22}^{\rm1.5PN_{\rm mem}}\right) \cr
	&+\frac{1}{c^5}\left( h_{22}^{\rm2.5PN_{\rm tail}}+h_{22}^{\rm2.5PN_{\rm mem}}+h_{22}^{\rm2.5PN_{\rm post-ad}}\right),
\end{align}



With this generic-orbit waveform information at hand, the aim of this section is to properly organize it so that it can be readily incorporated in the factorization scheme of  \dali{}~\cite{Chiaramello:2020ehz,Nagar:2020xsk,Nagar:2021gss}, in the most profitable way. To this end, for each $m\neq0$ mode we consider the following factorized structure
\begin{align}
	h_\lm = h_\lm^N  \hat{S}_{\rm eff}   \hat{h}_\lm^{\rm inst}  \hat{h}_\lm^{\rm QK},
\end{align}
where 
\begin{itemize}
	\item $h_\lm^N$ is the generic-planar-orbit Newtonian prefactor \cite{Chiaramello:2020ehz}, leading order term of $h_\lm^{\rm inst}$.
	\item $\hat{S}_{\rm eff}$ is the effective source term \cite{Damour:2008gu}, which is given by the $\mu$-rescaled effective EOB Hamiltonian $H_{\rm eff}$ when $\ell+m$ is even and by the Newton-normalized angular momentum $p_\varphi / r_\omega^2 \Omega $ when $\ell+m$ is odd, where $r_\omega$ is the modified EOB radius defined in Refs.~\cite{Damour:2006tr, Damour:2012ky}.
	\item $\hat{h}_\lm^{\rm inst}$ is a purely instantaneous, time-derivative dependent, PN factor. It is defined from $h_\lm^{\rm inst}$ by
	\begin{equation}
	\hat{h}_\lm^{\rm inst} \equiv T_{\rm 2.5PN} \left[\dfrac{h_\lm^{\rm inst}}{h_\lm^N \, \hat{S}_{\rm eff} } \right],
	\end{equation}
	where the operator $T_{\rm 2.5PN}$ applies to its argument a PN-type Taylor expansion up to 2.5PN, counting the PN orders with respect to the leading order of the full waveform, i.e. the Newtonian component of the $\l=m=2$ mode. This is exactly the instantaneous PN factor introduced in Ref,~\cite{Albanesi:2022xge}, now pushed to the 2.5PN order. Notice that the resulting structure is of the type ``1+PN corrections".
	\item $\hat{h}_\lm^{\rm QK}$ is the residual PN factor where all the other waveform contributions are collected, namely those derived in Sec.~\ref{sec:EOM_terms}. The formal definition of this factor is given by
	\begin{equation}
		\hat{h}_\lm^{\rm QK} \equiv T^{\rm EOM}_{\rm 2.5PN} \left[\dfrac{h_\lm}{h^{\rm inst}_\lm} \right],
	\end{equation}
	where the superscript ``EOM" on the Taylor series operator makes explicit that the expansion is taken using the PN-expanded EOM \emph{also} for the instantaneous part $h^{\rm inst}_\lm$ in the denominator. Again, it has a structure of the type ``1+PN corrections".
\end{itemize}

These last two PN factors are then further factorized by isolating their circular parts, namely
\begin{align}
	\hat{h}_\lm^{\rm X}& = \hat{h}_\lm^{\rm X_{\rm c}} \,  \hat{h}_\lm^{\rm X_{\rm nc}}, \quad \quad \hat{h}_\lm^{\rm X_{\rm nc}}\equiv T_{\rm 2.5PN} \left[\dfrac{	\hat{h}_\lm^{\rm X}}{\hat{h}_\lm^{\rm X_{\rm c}}} \right],
\end{align}
where $X=\{{\rm inst, QK}\}$.
Here the circular parts are computed by taking the circular limit, as clarified below, in the corresponding PN factors. For $\hat{h}_\lm^{\rm inst}$ this is simply realized by setting to zero $p_{r_{*}}$ and all the time derivatives of the EOB variables that appear therein except for $\Omega \equiv \dot{\varphi}$; notably this is done \emph{without} replacing the angular momentum $p_\varphi$ with its circular orbit expression in terms of $r$, as done in Ref.~\cite{Albanesi:2022xge}. In $\hat{h}_\lm^{\rm QK}$ we must take into account the expansion in eccentricity which underlies the waveform terms it incorporates. We do so as follows. First, we replace $p_\varphi$ with $\dot{p}_{r_*}$, by inverting perturbatively the EOB EOM of the latter, after it is expanded up to 2.5PN. Then, we take a simultaneous expansion in $p_{r_*}$ and $\dot{p}_{r_*}$ up to the sixth order, as we did for $\dot{r}_h$ and $k_p$ in Sec.~\ref{sec:EOM_terms}. At this point the circular part of this factor can be singled out by setting $p_{r_*}$ and $\dot{p}_{r_*}$ to zero. 

The resulting non-circular factors $\hat{h}_\lm^{\rm QK_{\rm nc}}$ are functions of $p_{r_*}$, $\dot{p}_{r_*}$  and $u$ that reduce to 1 when one takes either the Newtonian or the circular limit. We moreover split them into three distinct factors that separately collect tail, memory and post-adiabatic contributions, which give back the total factor $\hat{h}_\lm^{\rm QK_{\rm nc}}$ when multiplied together and expanded up to 2.5PN:
\begin{equation} \label{eq:splitQK}
    \hat{h}_\lm^{\rm QK_{\rm nc}} = \hat{h}_\lm^{\rm QK_{\rm nc,tail}}\hat{h}_\lm^{\rm QK_{\rm nc,mem}}\hat{h}_\lm^{\rm QK_{\rm nc,post-ad}}
\end{equation}

Globally, the factors $\hat{h}_\lm^{\rm inst_{\rm nc}}$ and the three factors in which $\hat{h}_\lm^{\rm QK_{\rm nc}}$ is split contain all the novel non-circular contributions to the waveform that we have computed in the previous section, in a form already set up for the inclusion in \dali{}; they are explicitly given in the supplementary Mathematica notebook. As for the circular factors $\hat{h}_\lm^{\rm inst_{\rm c}}$ and $\hat{h}_\lm^{\rm QK_{\rm c}}$, instead of keeping them as they are in the waveform model, we propose to replace them with the last avatars of the circular relativistic waveform factors $T_\lm e^{i \delta_\lm}$ and $(\rho_\lm)^\ell$ \cite{Damour:2008gu,Nagar:2020pcj}, used in all the previous iterations of the model.\footnote{Our intent here is to preserve as much as possible the great accuracy boasted by the native quasi-circular version of the model for the case of quasi-circular binary coalescences.} Here $T_\lm$ is a complex factor which resums infinite leading logarithms appearing in the tail part of the quasi-circular waveform,
 and is given by
	\begin{equation}
		T_\lm = \frac{\Gamma \big( \ell+1-2 i \hat{\hat{k}} \big) }{\Gamma\big(\ell+1\big)}\, e^{\pi \hat{\hat{k}}} \, e^{2 i \hat{\hat{k}} \log{(2 k r_0)}},
	\end{equation}
where $\hat{\hat{k}} \equiv G H_{\rm real} m \Omega$, in terms of the real EOB Hamiltonian $H_{\rm real}$, $k \equiv m \Omega$, and $r_0$ is a length scale introduced in the Blanchet-Damour waveform generation formalism \cite{Blanchet:1989ki}, fixed in this context to $r_0=2GM/\sqrt{e}$.
The other two quantities, $\delta_\lm$ and $\rho_\lm$, account respectively for the residual modulations to the phase and the amplitude of the spherical mode. They are given in terms of PN series which, being computed in the simplifying context of the quasi-circular approximation, span higher PN orders than the one reached by $\hat{h}_\lm^{\rm inst_{\rm c}}$ and $\hat{h}_\lm^{\rm QK_{\rm c}}$, especially in the test-mass ($\nu \to 0$) sector where they are pushed up to 5PN or 6PN accuracy, depending on the spherical mode; see e.g.~Ref.~\cite{Nagar:2020pcj}. Moreover, the behavior of the PN series in both $\delta_\lm$ and $\rho_\lm$ is tamed by specific Pad\'e resummations whose details can be found in Ref.~\cite{Nagar:2020pcj} and references therein.

To wrap up, the resulting factorized waveform model we propose for generic-planar-orbit black hole binaries reads
\begin{equation}
	h_\lm =  h_\lm^N  \hat{S}_{\rm eff} T_\lm e^{i \delta_\lm} (\rho_\lm)^\ell \hat{h}_\lm^{\rm inst_{\rm nc}}  \hat{h}_\lm^{\rm QK_{\rm nc}},
\end{equation}
with the non-circular factor $\hat{h}_\lm^{\rm QK_{\rm nc}}$ further split as in Eq.~\eqref{eq:splitQK}.
%
%
%
%
%
\section{Conclusions}
\label{sec:Conclusions}

In this work we have derived the 2.5PN accurate waveform components of each spherical mode of the waveform in EOB phase-space variables, in a form valid for binary systems moving along generic planar orbits. This extends the current knowledge of the PN-expanded EOB waveform with respect to previous works in two directions: (i) by including higher-order terms for all the waveform contributions that were already considered in the EOB literature, essentially instantaneous and tail terms, with the former provided either with explicit time derivatives or in the usual order-reduced form via the PN-expanded EOM; (ii) by introducing the 2.5PN-accurate EOB expression of the oscillatory memory terms, which so far have been missing in the EOB literature. We believe that our results will encourage and facilitate the inclusion of these neglected terms in EOB models, leading to a more comprehensive description of the gravitational wave signals radiated at infinity by non-circularized binaries. 

To further promote the application of the novel EOB waveform information we provide, we have have also computed associated non-circular factors that are suitably set up for being incorporated in the non-circular EOB model \dali{}. We defer to future work the assessment of the effective importance of these new corrections in \dali{}, as well as  the computation of the generic-planar-orbit EOB waveform at higher PN orders and the inclusion of extra spin-related corrections. 
\section{Acknowledgment}

G.G.~and M.O.~acknowledge financial support of the Ministero dell’Istruzione dell’Università e della Ricerca (MUR) through the program “Dipartimenti di Eccellenza 2018-2022” (Grant SUPER-C) and financial support from Fondo Ricerca di Base 2020 (MOSAICO) and 2021 (MEGA) of the University of Perugia.
The work of T.H. is supported in part by the project “Towards a deeper understanding of black holes with non-relativistic holography” of the Independent Research Fund Denmark (grant number DFF-6108-00340). 
G.G. and A.P.~thank the Niels Bohr Institute for hospitality at different stages of this project. T.H. thanks University of Perugia for hospitality.

\bibliography{refs,local}
\end{document}